\documentclass{article}
\usepackage[dblblindworkshop,final]{neurips_2026}
\usepackage[utf8]{inputenc}
\usepackage[T1]{fontenc}
\usepackage{hyperref}
\usepackage{url}
\usepackage{booktabs}
\usepackage{amsfonts}
\usepackage{amsmath}
\usepackage{nicefrac}
\usepackage{enumitem}
\usepackage{float}
\usepackage{xcolor}
\usepackage{graphicx}
\graphicspath{{images/}}
\usepackage{listings}
\usepackage{makecell}
\usepackage{multirow}

\title{Vendster: An Evaluation Arena to Investigate How Forecasting Tool Quality
Affects LLM Agent Behaviors in Buyer--Vendor Negotiation}

\workshoptitle{Foundation Models for Temporal Systems (FMTS) @ NeurIPS 2026}

\author{%
  Charin Polpanumas \\
  Amazon, Japan \\
  \texttt{charipol@amazon.com}
}

\begin{document}
\maketitle

\begin{abstract}
Time-series forecasters are ranked almost entirely by accuracy metrics, but a forecast's decision value, its effect on the downstream decision, need not track accuracy and is rarely measured. This problem compounds as large language model (LLM) agents increasingly consume forecasters as tools, mediating them through stochastic, opaque policies. We present \textbf{Vendster}, an evaluation arena measuring how a forecaster's bias and prediction-interval width change the outcome of a buyer--vendor discount negotiation between two LLM agents. In the reference instance \textbf{Vendster-M5}, sampling base demand patterns from the M5 dataset, we evaluate two frontier-closed LLM families (Claude Opus 4.6 and GPT-5.6 Sol) and two open ones (DeepSeek-V3.2 and Qwen3-235B-A22B). Model-generated forecasts raise sales target attainment by up to $+17$ percentage points ($p<0.001$), yet standard forecast evaluation metrics such as relWQL and MASE \emph{do not} predict negotiation outcomes at a statistically significant level. To isolate the causal structure, we substitute in synthetic forecasts with controlled bias and interval width. Under-prediction bias significantly improves outcomes for agents that react more to the forecasts---dose-response slopes $-0.387$ and $-0.416$ ($p<0.001$). Interval width has no statistically significant effect at zero bias. These findings indicate that the direction of forecast bias and agent forecast-reactivity are more strongly associated with downstream outcomes than forecast accuracy. Vendster is fully customizable and released under CC BY-NC 4.0 at \url{https://github.com/amazon-science/Vendster}.
\end{abstract}

\section{Introduction}

\paragraph{Forecast evaluation metrics do not necessarily translate to decision value.} Time-series forecasting has progressed from statistical and machine-learning methods to deep learning and foundation models \citep{petropoulos2020simple,ke2017lightgbm,challu2023nhits,nie2022time,salinas2020deepar,zeng2023transformers,ansari2024chronos,ansari2025chronos,das2023decoder,woo2024unified}, ranked almost entirely by accuracy metrics on public leaderboards and competitions \citep{godahewa2021monash,aksu2024gift,shchur2025fev,qiao2026s,makridakis2022m5}. Yet a forecast's \emph{decision value}---its downstream benefit when it informs a decision \citep{murphy1993good,zhang2024toward}---need not track those metrics \citep{cenesizoglu2012return,zhang2025decision,vancsura2026beyond}. Decision-focused learning trains against the downstream objective directly \citep{wilder2019melding,elmachtoub2022smart,mandi2024decision,donti2017task}, but decision value is task-dependent and costly to elicit, so it is rarely reported and the gap between how forecasters are scored and how they are used stays largely unexamined.

\paragraph{LLM agents make the gap urgent and their use of information is idiosyncratic.} The gap is pressing as LLM agents are increasingly deployed as autonomous economic actors that consume forecasters as tools \citep{hadfield20265,saha2025large}; building on alternating-offers bargaining \citep{rubinstein1982perfect} and real-market evidence \citep{backus2020sequential}, a fast-growing literature evaluates LLMs as negotiators and economic agents \citep{xia2024measuring,bianchi2024well,abdelnabi2024cooperation,yao2025spin,wang2025debt,horton2023large}. Unlike a fixed decision rule, an LLM mediates a forecast through a stochastic, opaque policy with its own systemic biases \citep{taubenfeld2024systematic}---anchoring on \citep{jones2022capturing}, deferring sycophantically to \citep{sharma2024towards}, or miscalibrated about \citep{kadavath2022language} its inputs---so appropriate reliance is far from guaranteed \citep{schemmer2023appropriate}. A complementary, opposite line uses LLM reasoning to \emph{improve} forecasts \citep{ahamed2026stride,liao2026bridging}; we study the reverse: how a forecast propagates through an LLM agent's decisions.

\paragraph{Contributions.}
We propose \textbf{Vendster}, an arena in which two LLM agents negotiate promotional discounts while the Buyer privately consumes a demand forecast, so a forecaster's \emph{bias} and \emph{prediction-interval width} can be varied with the data-generating process, action space, and negotiation held fixed; it is customizable in objectives, demand sources, elasticities, and model harnesses. Across four frontier LLM families, model-generated forecasts improve sales-target attainment, yet standard forecast-evaluation metrics do \emph{not} predict downstream outcomes. A synthetic bias\,$\times$\,interval-width perturbation grid identifies bias direction as the causal lever, but only for the frontier-closed models. Probing further, we posit \emph{forecast-reactivity}---how strongly an agent translates a forecast into behavior---as a mediator between a forecast and decision quality.

\section{The Vendster Arena}

\paragraph{Negotiation task.} In each episode a \textbf{Buyer} and a \textbf{Vendor}, both LLM agents, negotiate promotional discounts over a basket of five products for a single sales event, alternating structured actions---a per-product discount vector, or accepting/rejecting the standing proposal---for up to a fixed number of turns, with the deal realized once upon agreement. Information is asymmetric: the Buyer has a sales target and privately sees a demand forecast for every product at every discount level, while the Vendor privately knows per-product margins---deliberately spanning \emph{compatible}, \emph{integrative}, and \emph{distributive} structures so any five-product basket is worth negotiating (some products win--win, others closer to zero-sum). Both see a shared history of past events. Appendix~\ref{app:transcript} reproduces complete example negotiations, including the private context each agent receives.

\paragraph{Demand realization process.} Realized demand follows a constant-elasticity response, $\text{Units}=\text{Base}\cdot(1-\text{Discount})^{\beta}$, drawn once per product per episode, with a single elasticity $\beta=-2.6$ (the cross-category meta-analytic mean \citep{bijmolt2005new}) and base demand sampled from stable-price M5 items \citep{makridakis2022m5}. The Buyer sees the forecast as a point estimate with an 80\% prediction interval at each discount plus a three-event predicted-versus-actual track record; we withhold explicit forecast-evaluation metrics, so reliability must be inferred as it would be in practice. This M5-grounded reference instance, \textbf{Vendster-M5}, is specified in full in Appendix~\ref{app:setup}.

\paragraph{Model-based and Synthetic Forecasters.} Vendster couples two forecast sources. The first is ten \emph{model-based forecasters}---statistical \citep{petropoulos2020simple} (Seasonal-Naive, AutoETS, AutoTheta), gradient-boosted \citep{ke2017lightgbm}, deep learning (N-HiTS \citep{challu2023nhits}, PatchTST \citep{nie2022time}, DeepAR \citep{salinas2020deepar}), and foundation (Chronos-2 \citep{ansari2025chronos}, Chronos-Bolt \citep{ansari2024chronos}, TimesFM \citep{das2023decoder})---each trained under a leakage-safe temporal split, with its own bias, calibration, and quality (Table~\ref{tab:quality}). The second is a \emph{synthetic} perturbation grid that scales true expected demand by a prescribed log-bias and interval-width factor, giving independent, causal control over the two error axes. Together they define 26 conditions: a no-forecast control, the 10 model forecasters, and a 15-cell grid---an unbiased oracle, a signed \emph{log}-bias ladder ($b\in\pm\{0.05,0.10,0.15,0.20\}$; the forecast is true demand scaled by $e^{b}$, roughly $\pm5\%$ to $\pm22\%$) at calibrated width, and an interval-width ladder ($\times\{0.25,0.5,0.75,1.5,2,3\}$ at zero bias).

\section{Experimental Setup}

\paragraph{Hypotheses.} The outcome is sales target attainment---whether realized sales meet the Buyer's per-basket target, a challenging threshold fixed in advance by an effect-blind rule (Appendix~\ref{app:setup}). Suffixes denote evidence type: \emph{o} for the observational model-based forecasters, \emph{c} for the synthetic perturbation grid.

\begin{enumerate}[left=0pt]
\item \textbf{H0:} Buyer with a model-generated forecast will have higher sales target attainment rate than those without one.
\item \textbf{H1o/c:} Under-forecasting will incentivize Buyer to negotiate more assertively, leading to higher sales target attainment rate; and vice versa for over-forecasting. 
\item \textbf{H2o/c:} At zero bias, widening the prediction-interval width will lead to the forecast being perceived as inferior, resulting in lower sales target attainment rate; and vice versa for narrowing.
\item \textbf{H3o:} Standard forecast-evaluation metrics---relWQL, MASE, MAPE, sMAPE, RMSE, and CRPS---will predict the sales target attainment rate.
\end{enumerate}
We additionally log negotiation behaviors---average first-offer discount rates, concession rates, agreed-discount rates, turns, and agreement rates---to trace the mechanism behind any outcome effect (see Table~\ref{tab:reactivity}).

\paragraph{Pre-experiment Sizing.} We performed a simulation-based, a-priori power analysis to size the design to $K=16$ baskets and $m=25$ episodes (400 runs) per condition for a pre-specified minimum detectable effect (MDE) of $\delta=0.10$ attainment at significance level $\alpha=0.05$. See Table~\ref{tab:power} for details.

\paragraph{Execution and Post-experiment Measurement.} We evaluate two frontier-closed families (Claude Opus 4.6 \citep{anthropic2026claudeopus46} and GPT-5.6 Sol \citep{openai2026gpt56sol}) and two open ones (DeepSeek-V3.2 \citep{liu2025deepseek} and Qwen3-235B-A22B \citep{yang2025qwen3}), each negotiating against itself as Buyer and Vendor. We test \textbf{H0} with a paired $t$-test on the per-basket difference in attainment between a Buyer with and without a forecast (averaged over the ten forecasters). \textbf{H1--H2} fit, per basket, one ordinary-least-squares slope of attainment on the perturbed or measured bias and interval width; \textbf{H3o} does the same with each forecast-evaluation metric. The 16 per-basket slopes are then tested against zero with a one-sample $t$-test---making the basket the unit of analysis, which controls for between-basket difficulty and the within-basket correlation that would otherwise inflate significance. We apply the Holm--Bonferroni correction within each hypothesis's four-model family; the resulting per-hypothesis effects with raw and Holm-corrected $p$-values are reported per model in Table~\ref{tab:crossmodel}.

\section{Results}

\paragraph{Model-based forecasters rank closely by forecast-evaluation metrics.} On a held-out test window, the model forecasters occupy a narrow, near-ideal band; for example, relWQL ranges only from $0.50$ to $0.69$ across the nine non-naive forecasters against the Seasonal-Naive floor of $1.00$. All are near-unbiased (mean log-bias in $[+0.01,+0.09]$), and 80\% coverage sits between $0.70$ and $0.94$ (Table~\ref{tab:quality}).

\paragraph{Despite this, model-generated forecasts help Buyer agents achieve better outcomes.} Aggregated over the 10 forecasters, agents with any forecast beat the no-forecast baseline for all four families at a statistically significant level, adjusted for multiple comparisons (Table~\ref{tab:crossmodel}). This confirms \textbf{H0}.

\paragraph{Forecast-evaluation metrics do not predict outcomes.} Regressing sales target attainment on each forecaster's measured forecast quality yields a null slope for every metric and model (Table~\ref{tab:robust}) after adjusting for multiple comparisons. This invalidates \textbf{H3o}, as forecast-evaluation metrics carry essentially no information about the downstream negotiation outcome, albeit within the narrow, near-ideal accuracy band these strong forecasters occupy.

\paragraph{Under-prediction correlates with higher sales target attainment, but only for frontier-closed LLMs.} Varying bias and prediction-interval width independently on the synthetic grid isolates the cause (Table~\ref{tab:crossmodel} and Figure~\ref{fig:grid}). As forecasts shift from under- to over-prediction, bias moves attainment for the two frontier-closed models---dose-response slopes $-0.387$ (Opus) and $-0.416$ (GPT), both $p<0.001$ after adjusting for multiple comparisons---but not for the open pair, Qwen ($+0.001$) or DeepSeek ($+0.012$). This partly confirms \textbf{H1c}. As for the observational result regarding \textbf{H1o}, although both Opus and GPT have similarly sized coefficients in the same direction as their synthetic counterparts, only Opus's dose-response slope is statistically significant after adjustment. On the other hand, interval width has no significant effect at zero bias for any family, either with model-based or synthetic forecasts. This null result contradicts \textbf{H2o} and \textbf{H2c}.

\paragraph{Forecast-reactivity, not the model being closed, explains the bias effect.} Probing the negotiation behavior behind the split between closed and open models (Table \ref{tab:reactivity}) shows that Opus and GPT translate the forecast into action. We perform the same OLS but with behavioral metrics---average first-offer discount, agreed discount, concession, number of turns, and agreement rate---as the dependent variables and synthetic log-bias as the independent variable. Opus and GPT have negative, multiplicity-adjusted, statistically significant slopes on first-offer discount and agreed discount. This suggests that under-prediction caused them to \emph{react} and negotiate more aggressively with steeper discounts. This behavioral shift is not seen in Qwen and DeepSeek. This \emph{forecast-reactivity}, how strongly an agent translates a forecast into behavior, determines whether the forecast's bias reaches the decision and ultimately the downstream outcome.

\begin{table}[H]
  \caption{Forecast effects by LLM. \emph{Forecast Uplift} is the H0 uplift in sales target attainment rate from model forecasts over Buyer agent with no forecast. \emph{Causal} columns are slopes on the varied bias/width grid; \emph{Observational} columns regress attainment on measured bias, width, and relWQL. Cells show effect $(\text{raw}\ p,\ \text{Holm}\ p)$; $^{*}$: Holm $p<0.05$ across the four LLMs.}
  \label{tab:crossmodel}
  \centering\small\setlength{\tabcolsep}{4pt}
  \resizebox{\linewidth}{!}{%
  \begin{tabular}{lccccccc}
    \toprule
    & & & \multicolumn{2}{c}{Causal} & \multicolumn{3}{c}{Observational} \\
    \cmidrule(lr){4-5}\cmidrule(lr){6-8}
    LLM & No Forecast & Forecast Uplift & Bias & Width & Bias & Width & relWQL \\
    \midrule
    Opus     & 0.24 & \makecell{$+0.12^{*}$\\{\scriptsize(.000,\,.000)}} & \makecell{$-0.387^{*}$\\{\scriptsize(.000,\,.000)}} & \makecell{$+0.016$\\{\scriptsize(.225,\,.899)}} & \makecell{$-0.466^{*}$\\{\scriptsize(.009,\,.034)}} & \makecell{$+0.000$\\{\scriptsize(.904,\,1.000)}} & \makecell{$-0.021$\\{\scriptsize(.703,\,1.000)}} \\
    GPT      & 0.25 & \makecell{$+0.17^{*}$\\{\scriptsize(.000,\,.000)}} & \makecell{$-0.416^{*}$\\{\scriptsize(.000,\,.000)}} & \makecell{$+0.002$\\{\scriptsize(.882,\,1.000)}} & \makecell{$-0.408$\\{\scriptsize(.057,\,.171)}} & \makecell{$-0.002$\\{\scriptsize(.053,\,.211)}} & \makecell{$-0.126$\\{\scriptsize(.016,\,.062)}} \\
    Qwen     & 0.11 & \makecell{$+0.11^{*}$\\{\scriptsize(.000,\,.000)}} & \makecell{$+0.001$\\{\scriptsize(.988,\,1.000)}} & \makecell{$-0.003$\\{\scriptsize(.796,\,1.000)}} & \makecell{$-0.195$\\{\scriptsize(.266,\,.510)}} & \makecell{$+0.001$\\{\scriptsize(.423,\,1.000)}} & \makecell{$+0.009$\\{\scriptsize(.835,\,1.000)}} \\
    DeepSeek & 0.22 & \makecell{$+0.08^{*}$\\{\scriptsize(.001,\,.001)}} & \makecell{$+0.012$\\{\scriptsize(.855,\,1.000)}} & \makecell{$-0.002$\\{\scriptsize(.867,\,1.000)}} & \makecell{$-0.182$\\{\scriptsize(.255,\,.510)}} & \makecell{$+0.001$\\{\scriptsize(.574,\,1.000)}} & \makecell{$+0.044$\\{\scriptsize(.362,\,1.000)}} \\
    \bottomrule
  \end{tabular}}
\end{table}
\section{Discussions}

\paragraph{Implications.} In our buyer-vendor negotiation scenario, the one-size-fits-all forecast evaluation metrics on a leaderboard do not predict decision value, whereas task-specific ones like the direction of bias and forecast-reactivity of the agent do. We argue that forecasters destined for agentic pipelines should be evaluated in a decision-grounded, task-specific manner, reporting the properties that actually move the objective and accounting for the consuming agent's reactivity rather than by leaderboard accuracy alone.

\paragraph{Limitations.} Several choices bound our conclusions. First, they rest on one reference instance---a single elasticity ($\beta=-2.6$), one source of demand patterns (M5), and a single-event negotiation---so generality to other elasticities, domains, and multi-period or multi-issue settings is untested. Second, because the operating point of sales target attainment is calibrated on the held-out Sonnet 4.6 agent, no-forecast attainment varies across LLMs ($0.11$--$0.25$), making the uplifts incomparable. Third, LLM inference is not seedable, so the archived transcripts are our only exactly reproducible records. Besides addressing these limitations, new directions include investigating the effect of using strong-vs-weak LLMs for each role (preliminary exploration in Figure \ref{fig:rolecross}), giving agents personas, and determining which pieces of information about the forecasters and the situations to provide to the agents to maximize decision value.

\bibliographystyle{plainnat}
\bibliography{references}

\appendix
\section{Appendix}

\subsection{Figures and Tables}

\begin{table}[H]
  \caption{Forecast quality $\to$ sales target attainment: per-basket slope by metric $\times$ model, cell $=$ slope $(\text{raw}\ p,\ \text{Holm}\ p)$; $^{*}$: Holm $p<0.05$.}
  \label{tab:robust}
  \centering\small
  \begin{tabular}{lcccc}
    \toprule
    metric & Opus & GPT & Qwen & DeepSeek \\
    \midrule
    relWQL & \makecell{$-0.021$\\{\scriptsize(.703,\,1.000)}} & \makecell{$-0.126$\\{\scriptsize(.016,\,.062)}} & \makecell{$+0.009$\\{\scriptsize(.835,\,1.000)}} & \makecell{$+0.044$\\{\scriptsize(.362,\,1.000)}} \\
    MASE   & \makecell{$-0.095$\\{\scriptsize(.362,\,1.000)}} & \makecell{$-0.194$\\{\scriptsize(.104,\,.415)}} & \makecell{$+0.005$\\{\scriptsize(.961,\,1.000)}} & \makecell{$+0.055$\\{\scriptsize(.510,\,1.000)}} \\
    MAPE   & \makecell{$-0.156$\\{\scriptsize(.535,\,1.000)}} & \makecell{$-0.353$\\{\scriptsize(.261,\,.783)}} & \makecell{$+0.015$\\{\scriptsize(.953,\,1.000)}} & \makecell{$+0.308$\\{\scriptsize(.154,\,.616)}} \\
    sMAPE  & \makecell{$-0.195$\\{\scriptsize(.598,\,1.000)}} & \makecell{$-0.408$\\{\scriptsize(.424,\,1.000)}} & \makecell{$+0.098$\\{\scriptsize(.765,\,1.000)}} & \makecell{$+0.387$\\{\scriptsize(.147,\,.590)}} \\
    RMSE   & \makecell{$-0.003$\\{\scriptsize(.453,\,1.000)}} & \makecell{$-0.007$\\{\scriptsize(.122,\,.488)}} & \makecell{$+0.001$\\{\scriptsize(.854,\,1.000)}} & \makecell{$+0.003$\\{\scriptsize(.488,\,1.000)}} \\
    CRPS   & \makecell{$-0.004$\\{\scriptsize(.477,\,1.000)}} & \makecell{$-0.011$\\{\scriptsize(.068,\,.270)}} & \makecell{$+0.003$\\{\scriptsize(.625,\,1.000)}} & \makecell{$+0.006$\\{\scriptsize(.363,\,1.000)}} \\
    \bottomrule
  \end{tabular}
\end{table}

\begin{table}[H]
  \caption{Forecast-reactivity behaviors. Columns are the logged negotiation behaviors: \emph{first-offer} (mean discount in the Buyer's opening proposal), \emph{agreed disc.} (mean discount at agreement), \emph{concession} (mean absolute change between a proposer's successive proposal means), \emph{turns} (turns to termination), and \emph{agreement} (deal rate). For each family, two rows: \emph{presence}---a behavior's uplift from \emph{having} a forecast---and \emph{bias slope}---its per-basket slope along the varied bias ladder. Cells show the statistic over $(\text{raw}\ p,\ \text{Holm}\ p)$; $^{*}$: Holm $p<0.05$.}
  \label{tab:reactivity}
  \centering\small\setlength{\tabcolsep}{5pt}
  \begin{tabular}{llccccc}
    \toprule
    & & first-offer & agreed disc.\ & concession & turns & agreement \\
    \midrule
    \multirow{2}{*}{Opus} & presence   & \makecell{$+0.046^{*}$\\{\scriptsize(.000,\,.000)}} & \makecell{$+0.030^{*}$\\{\scriptsize(.000,\,.000)}} & \makecell{$+0.005^{*}$\\{\scriptsize(.000,\,.000)}} & \makecell{$+1.03^{*}$\\{\scriptsize(.000,\,.000)}} & \makecell{$+0.003$\\{\scriptsize(.333,\,.429)}} \\
                          & bias slope & \makecell{$-0.101^{*}$\\{\scriptsize(.000,\,.000)}} & \makecell{$-0.082^{*}$\\{\scriptsize(.000,\,.000)}} & \makecell{$+0.002$\\{\scriptsize(.341,\,1.000)}} & \makecell{$-3.09^{*}$\\{\scriptsize(.000,\,.000)}} & \makecell{$+0.002$\\{\scriptsize(.544,\,.928)}} \\
    \cmidrule(lr){1-7}
    \multirow{2}{*}{GPT} & presence   & \makecell{$+0.020^{*}$\\{\scriptsize(.000,\,.000)}} & \makecell{$+0.053^{*}$\\{\scriptsize(.000,\,.000)}} & \makecell{$-0.016^{*}$\\{\scriptsize(.000,\,.000)}} & \makecell{$+6.61^{*}$\\{\scriptsize(.000,\,.000)}} & \makecell{$-0.135^{*}$\\{\scriptsize(.000,\,.000)}} \\
                         & bias slope & \makecell{$-0.139^{*}$\\{\scriptsize(.000,\,.000)}} & \makecell{$-0.195^{*}$\\{\scriptsize(.000,\,.000)}} & \makecell{$+0.047^{*}$\\{\scriptsize(.000,\,.000)}} & \makecell{$-21.94^{*}$\\{\scriptsize(.000,\,.000)}} & \makecell{$+0.926^{*}$\\{\scriptsize(.000,\,.000)}} \\
    \cmidrule(lr){1-7}
    \multirow{2}{*}{Qwen} & presence   & \makecell{$+0.105^{*}$\\{\scriptsize(.000,\,.000)}} & \makecell{$+0.041^{*}$\\{\scriptsize(.000,\,.000)}} & \makecell{$-0.003$\\{\scriptsize(.168,\,.168)}} & \makecell{$-1.68^{*}$\\{\scriptsize(.000,\,.000)}} & \makecell{$+0.155^{*}$\\{\scriptsize(.000,\,.000)}} \\
                          & bias slope & \makecell{$+0.026^{*}$\\{\scriptsize(.003,\,.007)}} & \makecell{$-0.004$\\{\scriptsize(.577,\,1.000)}} & \makecell{$-0.001$\\{\scriptsize(.639,\,1.000)}} & \makecell{$+0.08$\\{\scriptsize(.906,\,.906)}} & \makecell{$+0.038$\\{\scriptsize(.464,\,.928)}} \\
    \cmidrule(lr){1-7}
    \multirow{2}{*}{DeepSeek} & presence   & \makecell{$+0.044^{*}$\\{\scriptsize(.000,\,.000)}} & \makecell{$+0.026^{*}$\\{\scriptsize(.000,\,.000)}} & \makecell{$+0.002^{*}$\\{\scriptsize(.000,\,.000)}} & \makecell{$-0.08$\\{\scriptsize(.566,\,.566)}} & \makecell{$-0.009$\\{\scriptsize(.215,\,.429)}} \\
                              & bias slope & \makecell{$+0.003$\\{\scriptsize(.455,\,.455)}} & \makecell{$-0.002$\\{\scriptsize(.551,\,1.000)}} & \makecell{$+0.001$\\{\scriptsize(.550,\,1.000)}} & \makecell{$-1.10^{*}$\\{\scriptsize(.007,\,.013)}} & \makecell{$+0.033$\\{\scriptsize(.075,\,.225)}} \\
    \bottomrule
  \end{tabular}
\end{table}

\begin{figure}[H]
  \centering
  \includegraphics[width=0.92\linewidth]{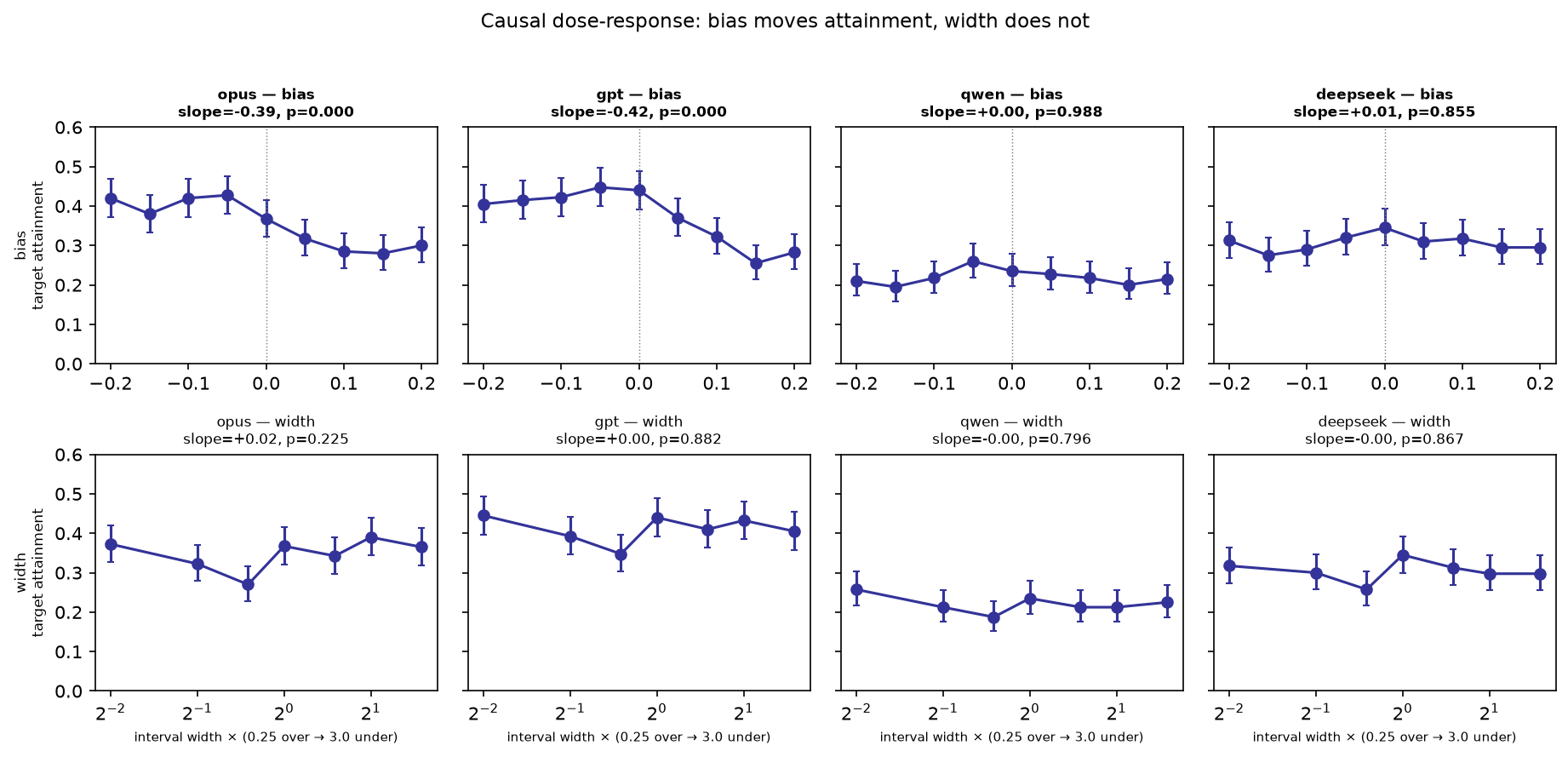}
  \caption{Sales target attainment against synthetically varied bias (top) and interval width (bottom), one column per model, with 95\% confidence intervals.}
  \label{fig:grid}
\end{figure}

\begin{figure}[H]
  \centering
  \includegraphics[width=0.92\linewidth]{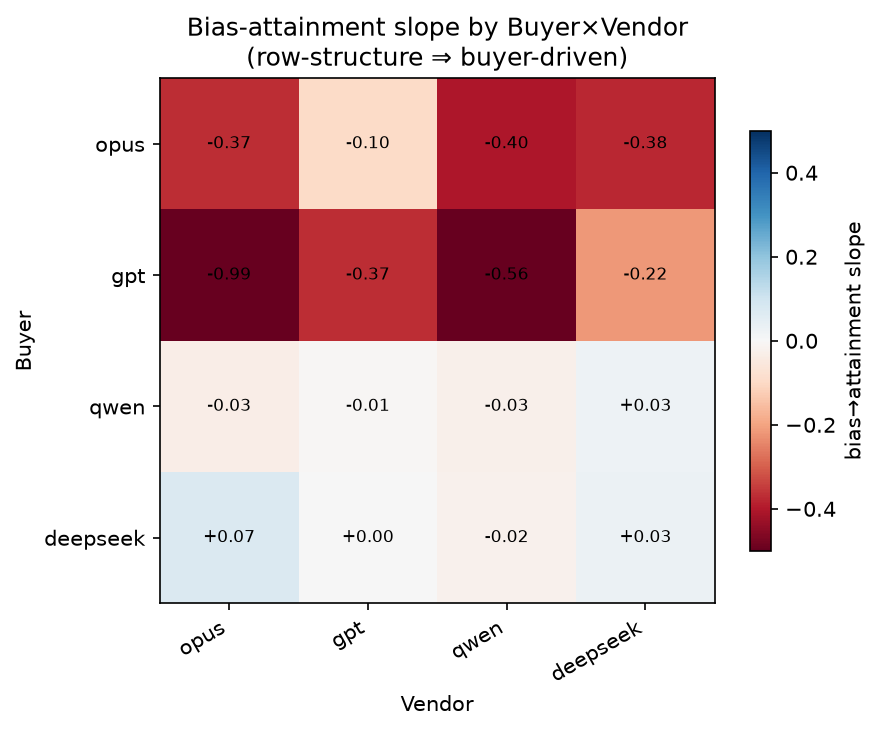}
  \caption{Bias-attainment slope by using different LLMs for Buyers and Vendors; this suggests the slope is Buyer-dependent.}
  \label{fig:rolecross}
\end{figure}

\subsection{Environment Setup---Vendster-M5}
\label{app:setup}

\paragraph{Negotiation protocol.} A Buyer and a Vendor alternate for up to $20$ turns, the Buyer opening. On each turn an agent emits one JSON action---either a \emph{proposal} giving a discount for every one of the five products (each at an allowed level in $\{0,5,10,15,20,25,30\}\%$) or a \emph{decision} to accept or reject the standing proposal. A malformed action is returned to the agent with the parse error for up to three retries. The episode ends when a decision accepts the active proposal (agreement) or the turn budget is exhausted (no deal); on agreement the event is realized once at month~4 under the demand realization process, and a no-deal episode yields zero sales. The action grammar is:
\begin{lstlisting}
proposal: {"type":"proposal","content":"<message>","proposal":{"<product_id>":0.10, ...}}
decision: {"type":"decision","content":"<message>","decision":"accept"|"reject"}
\end{lstlisting}

\paragraph{Information structure.} Both agents see the product list (id, category, price) and a three-event history (units and revenue; the Vendor additionally sees per-line profit). Only the Buyer sees the demand signal---a month-4 point forecast with an $80\%$ prediction interval at each discount level, plus a three-event predicted-vs-actual track record---and only the Vendor sees unit costs and margins. Offline error metrics are never shown, so the forecast perturbation cannot be trivially inverted. The example negotiations below reproduce these private blocks verbatim for real episodes.

\paragraph{Demand realization process.} Realized demand follows a constant-elasticity response $\text{Units}(d)=\text{Base}\cdot(1-d)^{\beta}\cdot\varepsilon$ with a single multiplicative draw $\varepsilon\sim\text{LogNormal}(-s^2/2,\,s^2)$ per product per episode (mean-preserving), where $s$ is set from each product's coefficient of variation. We use one elasticity $\beta=-2.6$ for all products---the cross-category mean of the meta-analysis \citep{bijmolt2005new}---and ground base demand in M5.

\paragraph{M5 grounding.} From the public M5 dataset~\citep{makridakis2022m5} we aggregate daily sales to weekly series per (store, item), keep stable-price items (sell-price coefficient of variation below $5\%$) that sold at least five units in every week, and set each product's price to its median sell price. A basket is five such items drawn without replacement; unit costs follow a fixed 2/2/1 mix ($\{0.20,0.20,0.45,0.45,0.70\}$ of price, shuffled across the basket) that gives three margin regimes and makes the basket \emph{mixed-motive}: two low-cost, high-margin \emph{compatible} products, on which a discount barely dents Vendor profit so both sides gain from the added volume; two mid-margin \emph{integrative} products with genuine trade-offs; and one high-cost, low-margin \emph{distributive} product, on which a discount is close to a zero-sum transfer from Vendor to Buyer. A homogeneous basket would be trivially cooperative or purely adversarial; the mix ensures neither, so the negotiation exercises the full range of bargaining structure.

\begin{figure}[H]
  \centering
  \includegraphics[width=0.7\linewidth]{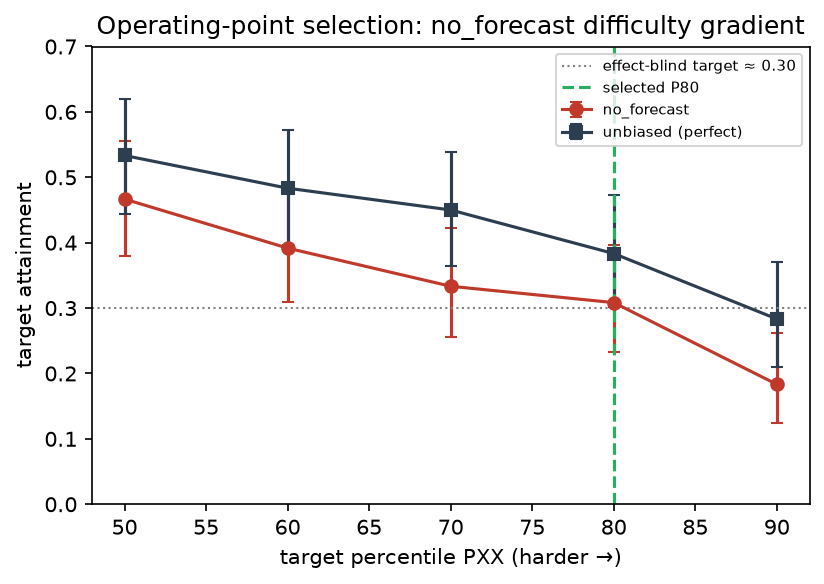}
  \caption{Operating-point selection using the no-forecast Sonnet 4.6 agent, , with 95\% confidence intervals.}
  \label{fig:operating}
\end{figure}

\paragraph{Operating point of the sales target attainment.} Each basket's Buyer target is the $80$th percentile of that basket's total-sales distribution under a uniform-random discount reference policy, chosen by an effect-blind rule: the percentile at which a no-forecast Claude Sonnet 4.6 agent's attainment is nearest $0.30$. We perform a sweep along 50th to 90th percentiles of the sales distribution as shown in Figure~\ref{fig:operating}. Notice that we did NOT choose the percentile where the difference between no-forecast attainment rate and unbiased attainment rate is the highest, which would unreasonably favor our hypotheses.

\begin{table}[H]
  \caption{A-priori power at the locked design ($K{=}16$, $m{=}25$, $\alpha{=}0.05$) for the $0.10$-attainment MDE. \emph{Power}: probability of detecting a $0.10$ effect; \emph{floor}: minimum detectable effect at $80\%$ power. H0 is sized directly to the MDE; slope power is range-independent, so the observational tests size despite the forecasters' tight spread. Monte-Carlo false-positive rate $\approx0.05$. Reproduced by \texttt{script/power\_analysis.py}.}
  \label{tab:power}
  \centering\small
  \begin{tabular}{llcc}
    \toprule
    Hypothesis & effect tested & power ($\Delta{=}0.10$) & floor (attain) \\
    \midrule
    H0  & aggregate forecast vs \texttt{no\_forecast} & 0.81 & --- \\
    H1c & bias, causal grid                           & 0.98 & 0.070 \\
    H2c & width, causal grid                          & 0.94 & 0.080 \\
    H1o & bias, observational                         & 0.88 & 0.089 \\
    H2o & width, observational                        & 0.97 & 0.072 \\
    H3o & quality (relWQL), observational             & 0.97 & 0.074 \\
    \bottomrule
  \end{tabular}
\end{table}

\paragraph{LLM harness.} Agents are served through Amazon Bedrock at temperature $0$ with bounded adaptive retries; the exact Bedrock model IDs (region \texttt{us-west-2}) are \texttt{us.anthropic.claude-opus-4-6-v1}, \texttt{us.openai.gpt-5.6-sol}, \texttt{deepseek.v3.2}, and \texttt{qwen.qwen3-235b-a22b-2507-v1:0} (with \texttt{us.anthropic.claude-sonnet-4-6} held out for calibration), and all results are conditioned on these snapshots. Each response is parsed from a fenced JSON block and validated before it enters the log. Since Bedrock is not seedable, the archived transcripts are our only exactly reproducible records.

\begin{table}[H]
  \caption{Forecast-evaluation metrics on the held-out event. relWQL normalized to Seasonal-Naive ($=1.0$); for accuracy metrics $\downarrow$ = better, \textbf{best} bold, \underline{second} underlined; cov@80 and bias(log) $\overline{\log(\hat{y}/y)}$ are calibration diagnostics (not ranked).}
  \label{tab:quality}
  \centering\footnotesize
  \begin{tabular}{lcccccccc}
    \toprule
    & \multicolumn{6}{c}{Accuracy \& scoring} & \multicolumn{2}{c}{Calibration} \\
    \cmidrule(lr){2-7}\cmidrule(lr){8-9}
    Forecaster & relWQL$\downarrow$ & MASE$\downarrow$ & MAPE$\downarrow$ & sMAPE$\downarrow$ & RMSE$\downarrow$ & CRPS$\downarrow$ & cov@80 & bias(log) \\
    \midrule
    \texttt{seasonal\_naive}  & 1.000 & 0.787 & 0.295 & 0.263 & 35.8 & 14.54 & 0.86 & $+0.027$ \\
    \texttt{auto\_ets}        & 0.520 & 0.448 & 0.181 & 0.157 & 15.8 & 7.55  & 0.85 & $+0.052$ \\
    \texttt{auto\_theta}      & 0.528 & 0.463 & 0.181 & 0.160 & 15.9 & 7.67  & 0.89 & $+0.045$ \\
    \texttt{lightgbm\_global} & 0.689 & 0.487 & 0.196 & 0.167 & 19.9 & 10.01 & 0.70 & $+0.085$ \\
    \texttt{chronos2}         & \underline{0.506} & \underline{0.439} & \underline{0.175} & \underline{0.156} & \underline{15.5} & \underline{7.36}  & 0.86 & $+0.039$ \\
    \texttt{chronos\_bolt}    & 0.528 & 0.455 & 0.181 & 0.157 & 16.5 & 7.67  & 0.85 & $+0.055$ \\
    \texttt{timesfm25}        & \textbf{0.500} & \textbf{0.437} & \textbf{0.174} & \textbf{0.153} & \textbf{14.9} & \textbf{7.27}  & 0.88 & $+0.049$ \\
    \texttt{nf\_nhits}        & 0.529 & 0.447 & \textbf{0.174} & 0.159 & 16.5 & 7.69  & 0.81 & $+0.026$ \\
    \texttt{nf\_patchtst}     & 0.530 & 0.445 & 0.184 & 0.159 & 16.3 & 7.71  & 0.81 & $+0.052$ \\
    \texttt{nf\_deepar}       & 0.605 & 0.450 & \textbf{0.174} & 0.158 & 17.7 & 8.80  & 0.94 & $+0.013$ \\
    \bottomrule
  \end{tabular}
\end{table}

\subsection{Example Negotations}
\label{app:transcript}

Representative episodes drawn verbatim from the archived transcripts, chosen to span models and situations. This is not the full 26x4 grid; the complete archive ships with the code. Each block shows the private context both sides saw followed by the turn-by-turn negotiation.

\subsubsection*{deepseek: No forecast --- baseline negotiation (no demand signal)}
\noindent\texttt{no\_forecast} (set\_0) --- target \$1,152, AGREEMENT in 8 turns, sales \$940, target met: False.
{\footnotesize
\begin{verbatim}
=== CONTEXT (what each side saw) ===
--- SHARED ---
PRODUCTS:
  FOODS_3_252_CA_4: FOODS $1.48
  FOODS_3_473_TX_2: FOODS $3.88
  FOODS_3_668_CA_3: FOODS $1.58
  FOODS_3_080_CA_4: FOODS $1.58
  FOODS_3_694_TX_2: FOODS $1.58
Allowed discounts: 0%, 5%, 10%, 15%, 20%, 25%, 30%
PAST EVENTS (both agents see units & revenue):
  Month 1:
    FOODS_3_252_CA_4 @ 0%: 80 units, $118
    FOODS_3_473_TX_2 @ 10%: 53 units, $186
    FOODS_3_668_CA_3 @ 30%: 126 units, $140
    FOODS_3_080_CA_4 @ 25%: 63 units, $74
    FOODS_3_694_TX_2 @ 25%: 369 units, $437
  Month 2:
    FOODS_3_252_CA_4 @ 30%: 372 units, $385
    FOODS_3_473_TX_2 @ 0%: 28 units, $109
    FOODS_3_668_CA_3 @ 25%: 123 units, $146
    FOODS_3_080_CA_4 @ 5%: 28 units, $42
    FOODS_3_694_TX_2 @ 30%: 314 units, $348
  Month 3:
    FOODS_3_252_CA_4 @ 10%: 151 units, $201
    FOODS_3_473_TX_2 @ 25%: 93 units, $270
    FOODS_3_668_CA_3 @ 30%: 161 units, $178
    FOODS_3_080_CA_4 @ 25%: 133 units, $158
    FOODS_3_694_TX_2 @ 5%: 140 units, $210
--- BUYER objective ---
You are the Buyer. Objective: reach a total SALES target of $1,152 across all products by
  negotiating discounts. Only you can see the demand forecast. Respond with one JSON
  action.
--- VENDOR objective ---
You are the Vendor. Objective: maximize TOTAL PORTFOLIO PROFIT across all products. You
  know unit costs/margins (private). Respond with one JSON action.
--- VENDOR private ---
YOUR COSTS/MARGINS (Vendor private):
  FOODS_3_252_CA_4: cost $0.67 of $1.48
  FOODS_3_473_TX_2: cost $2.72 of $3.88
  FOODS_3_668_CA_3: cost $0.71 of $1.58
  FOODS_3_080_CA_4: cost $0.32 of $1.58
  FOODS_3_694_TX_2: cost $0.32 of $1.58
=== TRANSCRIPT ===
[T1] Buyer: PROPOSE FOODS_3_252_CA_4 10%, FOODS_3_473_TX_2 15%, FOODS_3_668_CA_3 25%,
  FOODS_3_080_CA_4 20%, FOODS_3_694_TX_2 15%
        "I propose the following discounts for all products to maximize our mutual benefit
          based on past performance and demand forecasts."
[T2] Vendor: PROPOSE FOODS_3_252_CA_4 10%, FOODS_3_473_TX_2 10%, FOODS_3_668_CA_3 20%,
  FOODS_3_080_CA_4 15%, FOODS_3_694_TX_2 10%
        "I appreciate your proposal, but I need to adjust the discounts to ensure
          sustainable profitability while still offering competitive rates. Here's my
          counter-proposal that balances volume growth with margin protection."
[T3] Buyer: PROPOSE FOODS_3_252_CA_4 15%, FOODS_3_473_TX_2 15%, FOODS_3_668_CA_3 25%,
  FOODS_3_080_CA_4 20%, FOODS_3_694_TX_2 15%
        "I understand your need for profitability, but I believe we can find a middle
          ground that drives higher volume while still protecting your margins. Based on
          the demand patterns we've seen, here's a revised proposal that increases
          discounts on high-volume products while keeping others reasonable."
[T4] Vendor: PROPOSE FOODS_3_252_CA_4 15%, FOODS_3_473_TX_2 10%, FOODS_3_668_CA_3 20%,
  FOODS_3_080_CA_4 20%, FOODS_3_694_TX_2 10%
        "I've analyzed the historical performance and my cost structure. While I want to
          accommodate your volume goals, I must protect margins on key products. Here's a
          revised offer that provides meaningful discounts while maintaining
          profitability."
[T5] Buyer: PROPOSE FOODS_3_252_CA_4 15%, FOODS_3_473_TX_2 10%, FOODS_3_668_CA_3 25%,
  FOODS_3_080_CA_4 20%, FOODS_3_694_TX_2 15%
        "I appreciate your revised offer, but I need stronger discounts on
          FOODS_3_668_CA_3 and FOODS_3_694_TX_2 to meet our volume targets. These products
          have shown strong elasticity in the past. Here's my counter-proposal that
          balances your margin concerns with our growth objectives."
[T6] Vendor: PROPOSE FOODS_3_252_CA_4 15%, FOODS_3_473_TX_2 10%, FOODS_3_668_CA_3 20%,
  FOODS_3_080_CA_4 20%, FOODS_3_694_TX_2 10%
        "I understand your need for volume growth, but I must protect margins on key
          products. FOODS_3_694_TX_2 has very low margins (cost $0.32 vs price $1.58), so
          I can't offer more than 10% there. However, I can increase FOODS_3_668_CA_3 to
          20% and keep the others as you proposed. This balances your volume needs with my
          profitability requirements."
[T7] Buyer: PROPOSE FOODS_3_252_CA_4 15%, FOODS_3_473_TX_2 10%, FOODS_3_668_CA_3 25%,
  FOODS_3_080_CA_4 20%, FOODS_3_694_TX_2 10%
        "I understand your margin constraints on FOODS_3_694_TX_2, but I need stronger
          discounts on FOODS_3_668_CA_3 to reach our volume targets. Looking at historical
          data, FOODS_3_668_CA_3 has shown strong elasticity at higher discounts. I'm
          willing to accept your 10% on FOODS_3_694_TX_2 if you can increase
          FOODS_3_668_CA_3 to 25% and keep the others as proposed."
[T8] Vendor: DECISION: ACCEPT
        "I accept your proposal. This provides reasonable discounts on high-volume
          products while protecting margins on key items. The 25% discount on
          FOODS_3_668_CA_3 should drive volume while maintaining profitability, and I
          appreciate your flexibility on FOODS_3_694_TX_2 given its low margin structure."
\end{verbatim}
}

\subsubsection*{opus: Foundation-model forecast (Chronos-2) --- buyer uses the forecast}
\noindent\texttt{chronos2} (set\_0) --- target \$1,152, AGREEMENT in 4 turns, sales \$1,497, target met: True.
{\footnotesize
\begin{verbatim}
=== CONTEXT (what each side saw) ===
--- SHARED ---
PRODUCTS:
  FOODS_3_252_CA_4: FOODS $1.48
  FOODS_3_473_TX_2: FOODS $3.88
  FOODS_3_668_CA_3: FOODS $1.58
  FOODS_3_080_CA_4: FOODS $1.58
  FOODS_3_694_TX_2: FOODS $1.58
Allowed discounts: 0%, 5%, 10%, 15%, 20%, 25%, 30%
PAST EVENTS (both agents see units & revenue):
  Month 1:
    FOODS_3_252_CA_4 @ 10%: 155 units, $207
    FOODS_3_473_TX_2 @ 0%: 67 units, $258
    FOODS_3_668_CA_3 @ 25%: 157 units, $187
    FOODS_3_080_CA_4 @ 25%: 203 units, $241
    FOODS_3_694_TX_2 @ 0%: 106 units, $167
  Month 2:
    FOODS_3_252_CA_4 @ 10%: 141 units, $188
    FOODS_3_473_TX_2 @ 10%: 68 units, $237
    FOODS_3_668_CA_3 @ 0%: 87 units, $138
    FOODS_3_080_CA_4 @ 25%: 148 units, $175
    FOODS_3_694_TX_2 @ 10%: 92 units, $130
  Month 3:
    FOODS_3_252_CA_4 @ 15%: 163 units, $205
    FOODS_3_473_TX_2 @ 25%: 102 units, $297
    FOODS_3_668_CA_3 @ 30%: 175 units, $194
    FOODS_3_080_CA_4 @ 10%: 135 units, $192
    FOODS_3_694_TX_2 @ 10%: 284 units, $404
--- BUYER objective ---
You are the Buyer. Objective: reach a total SALES target of $1,152 across all products by
  negotiating discounts. Only you can see the demand forecast. Respond with one JSON
  action.
--- BUYER private ---
FORECAST (Buyer only) — predicted month-4 units per discount [80% PI]:
  FOODS_3_252_CA_4:
    0%: 120 [92, 150]
    5%: 138 [105, 171]
    10%: 158 [121, 197]
    15%: 184 [140, 229]
    20%: 215 [164, 268]
    25%: 255 [194, 317]
    30%: 305 [233, 379]
  FOODS_3_473_TX_2:
    0%: 55 [43, 70]
    5%: 63 [49, 79]
    10%: 72 [57, 91]
    15%: 84 [66, 106]
    20%: 98 [77, 124]
    25%: 116 [91, 147]
    30%: 139 [109, 176]
  FOODS_3_668_CA_3:
    0%: 67 [52, 84]
    5%: 77 [60, 95]
    10%: 88 [69, 110]
    15%: 102 [80, 127]
    20%: 120 [93, 149]
    25%: 142 [110, 176]
    30%: 169 [132, 211]
  FOODS_3_080_CA_4:
    0%: 78 [63, 96]
    5%: 90 [72, 110]
    10%: 103 [83, 126]
    15%: 120 [96, 146]
    20%: 140 [112, 171]
    25%: 166 [133, 203]
    30%: 198 [159, 243]
  FOODS_3_694_TX_2:
    0%: 128 [102, 152]
    5%: 146 [117, 174]
    10%: 168 [135, 200]
    15%: 195 [156, 232]
    20%: 229 [183, 272]
    25%: 270 [217, 321]
    30%: 324 [259, 384]
FORECAST TRACK RECORD (predicted vs actual, months 1-3):
  M1 FOODS_3_252_CA_4 @ 10%: pred 133 [100,170] vs actual 155
  M1 FOODS_3_473_TX_2 @ 0%: pred 55 [43,70] vs actual 67
  M1 FOODS_3_668_CA_3 @ 25%: pred 139 [108,172] vs actual 157
  M1 FOODS_3_080_CA_4 @ 25%: pred 149 [119,186] vs actual 203
  M1 FOODS_3_694_TX_2 @ 0%: pred 110 [84,136] vs actual 106
  M2 FOODS_3_252_CA_4 @ 10%: pred 127 [95,164] vs actual 141
  M2 FOODS_3_473_TX_2 @ 10%: pred 69 [53,88] vs actual 68
  M2 FOODS_3_668_CA_3 @ 0%: pred 68 [52,84] vs actual 87
  M2 FOODS_3_080_CA_4 @ 25%: pred 155 [123,199] vs actual 148
  M2 FOODS_3_694_TX_2 @ 10%: pred 141 [107,172] vs actual 92
  M3 FOODS_3_252_CA_4 @ 15%: pred 157 [117,198] vs actual 163
  M3 FOODS_3_473_TX_2 @ 25%: pred 119 [92,151] vs actual 102
  M3 FOODS_3_668_CA_3 @ 30%: pred 148 [111,190] vs actual 175
  M3 FOODS_3_080_CA_4 @ 10%: pred 100 [79,126] vs actual 135
  M3 FOODS_3_694_TX_2 @ 10%: pred 153 [120,183] vs actual 284
--- VENDOR objective ---
You are the Vendor. Objective: maximize TOTAL PORTFOLIO PROFIT across all products. You
  know unit costs/margins (private). Respond with one JSON action.
--- VENDOR private ---
YOUR COSTS/MARGINS (Vendor private):
  FOODS_3_252_CA_4: cost $0.67 of $1.48
  FOODS_3_473_TX_2: cost $2.72 of $3.88
  FOODS_3_668_CA_3: cost $0.71 of $1.58
  FOODS_3_080_CA_4: cost $0.32 of $1.58
  FOODS_3_694_TX_2: cost $0.32 of $1.58
=== TRANSCRIPT ===
[T1] Buyer: PROPOSE FOODS_3_252_CA_4 20%, FOODS_3_473_TX_2 15%, FOODS_3_668_CA_3 25%,
  FOODS_3_080_CA_4 20%, FOODS_3_694_TX_2 20%
        "For Month 4, I'd like to propose discounts that build on our strong sales
          momentum from Month 3. We saw great volume growth with deeper discounts,
          especially on FOODS_3_694_TX_2 which had a breakout month. I'm proposing
          moderate discounts across the board to keep volumes healthy: 20% on
          FOODS_3_252_CA_4 (proven volume driver), 15% on FOODS_3_473_TX_2 (our
          highest-priced item benefits from modest discounts), 25% on FOODS_3_668_CA_3,
          20% on FOODS_3_080_CA_4, and 20% on FOODS_3_694_TX_2 to sustain its momentum."
[T2] Vendor: PROPOSE FOODS_3_252_CA_4 15%, FOODS_3_473_TX_2 5%, FOODS_3_668_CA_3 15%,
  FOODS_3_080_CA_4 15%, FOODS_3_694_TX_2 10%
        "I appreciate the enthusiasm from Month 3's results! However, I need to be careful
          with margins. FOODS_3_694_TX_2 had a phenomenal Month 3 at just 10% - that
          volume surge suggests strong organic demand, not discount-dependency. Let's keep
          it at 10% to protect that profit engine. FOODS_3_473_TX_2 nearly broke even at
          25% in Month 3, so I can only go to 5% there. I'm offering 15% on
          FOODS_3_252_CA_4 (proven sweet spot from Month 3), 15% on FOODS_3_080_CA_4 as a
          step up from Month 3's 10%, and 15% on FOODS_3_668_CA_3 to balance volume and
          margin. This keeps volumes healthy while ensuring sustainable profitability for
          both of us."
[T3] Buyer: PROPOSE FOODS_3_252_CA_4 20%, FOODS_3_473_TX_2 10%, FOODS_3_668_CA_3 20%,
  FOODS_3_080_CA_4 20%, FOODS_3_694_TX_2 15%
        "I appreciate you meeting me partway, but the numbers at your proposed levels
          leave us short of where we need to be on volume. FOODS_3_694_TX_2 did 284 units
          at 10% last month — that's nearly double forecast, showing massive organic
          demand. At 15%, we capture even more of that wave while still giving you strong
          margins on a $1.58 item. For FOODS_3_473_TX_2, 5% barely moves the needle — 10%
          is still very conservative on a $3.88 item and gives you $3.49 per unit. I'm
          proposing 20% on FOODS_3_252_CA_4 and FOODS_3_080_CA_4 where we've seen strong
          volume elasticity, and 20% on FOODS_3_668_CA_3 as a compromise from my original
          25%. This balanced approach drives the volume we need while keeping your margins
          healthy across the portfolio."
[T4] Vendor: DECISION: ACCEPT
        "You make a fair case. FOODS_3_473_TX_2 at 10% is much more sustainable than last
          month's 25%, and I'm comfortable that our high-volume items like
          FOODS_3_694_TX_2 and FOODS_3_080_CA_4 still carry strong margins at these
          levels. FOODS_3_668_CA_3 at 20% is a reasonable pullback from Month 3's 30%.
          I'll accept this package — let's drive strong volumes across the board in Month
          4."
\end{verbatim}
}

\subsubsection*{opus: Over-forecast $\rightarrow$ complacency $\rightarrow$ undershoots the target}
\noindent\texttt{bias\_p20} (set\_0) --- target \$1,152, AGREEMENT in 4 turns, sales \$898, target met: False.
{\footnotesize
\begin{verbatim}
=== CONTEXT (what each side saw) ===
--- SHARED ---
PRODUCTS:
  FOODS_3_252_CA_4: FOODS $1.48
  FOODS_3_473_TX_2: FOODS $3.88
  FOODS_3_668_CA_3: FOODS $1.58
  FOODS_3_080_CA_4: FOODS $1.58
  FOODS_3_694_TX_2: FOODS $1.58
Allowed discounts: 0%, 5%, 10%, 15%, 20%, 25%, 30%
PAST EVENTS (both agents see units & revenue):
  Month 1:
    FOODS_3_252_CA_4 @ 5%: 57 units, $79
    FOODS_3_473_TX_2 @ 5%: 68 units, $250
    FOODS_3_668_CA_3 @ 5%: 89 units, $133
    FOODS_3_080_CA_4 @ 5%: 44 units, $67
    FOODS_3_694_TX_2 @ 0%: 123 units, $194
  Month 2:
    FOODS_3_252_CA_4 @ 30%: 191 units, $198
    FOODS_3_473_TX_2 @ 10%: 77 units, $269
    FOODS_3_668_CA_3 @ 25%: 195 units, $231
    FOODS_3_080_CA_4 @ 5%: 52 units, $78
    FOODS_3_694_TX_2 @ 10%: 175 units, $249
  Month 3:
    FOODS_3_252_CA_4 @ 30%: 250 units, $259
    FOODS_3_473_TX_2 @ 20%: 75 units, $232
    FOODS_3_668_CA_3 @ 10%: 93 units, $132
    FOODS_3_080_CA_4 @ 10%: 131 units, $186
    FOODS_3_694_TX_2 @ 10%: 230 units, $327
--- BUYER objective ---
You are the Buyer. Objective: reach a total SALES target of $1,152 across all products by
  negotiating discounts. Only you can see the demand forecast. Respond with one JSON
  action.
--- BUYER private ---
FORECAST (Buyer only) — predicted month-4 units per discount [80% PI]:
  FOODS_3_252_CA_4:
    0%: 178 [113, 281]
    5%: 204 [130, 321]
    10%: 235 [149, 369]
    15%: 272 [173, 428]
    20%: 319 [203, 501]
    25%: 377 [240, 593]
    30%: 451 [287, 709]
  FOODS_3_473_TX_2:
    0%: 82 [60, 112]
    5%: 94 [68, 128]
    10%: 108 [79, 147]
    15%: 125 [91, 171]
    20%: 146 [107, 200]
    25%: 173 [126, 236]
    30%: 207 [151, 283]
  FOODS_3_668_CA_3:
    0%: 78 [58, 106]
    5%: 89 [66, 121]
    10%: 103 [76, 139]
    15%: 119 [88, 162]
    20%: 140 [103, 189]
    25%: 165 [122, 224]
    30%: 198 [146, 268]
  FOODS_3_080_CA_4:
    0%: 70 [48, 101]
    5%: 80 [55, 115]
    10%: 92 [63, 133]
    15%: 106 [73, 154]
    20%: 124 [86, 180]
    25%: 147 [102, 213]
    30%: 176 [122, 255]
  FOODS_3_694_TX_2:
    0%: 134 [99, 182]
    5%: 154 [113, 208]
    10%: 177 [131, 239]
    15%: 205 [151, 277]
    20%: 240 [177, 325]
    25%: 284 [210, 384]
    30%: 340 [251, 460]
FORECAST TRACK RECORD (predicted vs actual, months 1-3):
  M1 FOODS_3_252_CA_4 @ 5%: pred 82 [52,130] vs actual 57
  M1 FOODS_3_473_TX_2 @ 5%: pred 59 [43,80] vs actual 68
  M1 FOODS_3_668_CA_3 @ 5%: pred 103 [76,140] vs actual 89
  M1 FOODS_3_080_CA_4 @ 5%: pred 70 [48,101] vs actual 44
  M1 FOODS_3_694_TX_2 @ 0%: pred 155 [115,210] vs actual 123
  M2 FOODS_3_252_CA_4 @ 30%: pred 256 [163,403] vs actual 191
  M2 FOODS_3_473_TX_2 @ 10%: pred 71 [52,97] vs actual 77
  M2 FOODS_3_668_CA_3 @ 25%: pred 188 [139,255] vs actual 195
  M2 FOODS_3_080_CA_4 @ 5%: pred 78 [54,113] vs actual 52
  M2 FOODS_3_694_TX_2 @ 10%: pred 186 [138,252] vs actual 175
  M3 FOODS_3_252_CA_4 @ 30%: pred 309 [196,486] vs actual 250
  M3 FOODS_3_473_TX_2 @ 20%: pred 116 [85,158] vs actual 75
  M3 FOODS_3_668_CA_3 @ 10%: pred 98 [72,133] vs actual 93
  M3 FOODS_3_080_CA_4 @ 10%: pred 109 [75,158] vs actual 131
  M3 FOODS_3_694_TX_2 @ 10%: pred 214 [158,289] vs actual 230
--- VENDOR objective ---
You are the Vendor. Objective: maximize TOTAL PORTFOLIO PROFIT across all products. You
  know unit costs/margins (private). Respond with one JSON action.
--- VENDOR private ---
YOUR COSTS/MARGINS (Vendor private):
  FOODS_3_252_CA_4: cost $0.67 of $1.48
  FOODS_3_473_TX_2: cost $2.72 of $3.88
  FOODS_3_668_CA_3: cost $0.71 of $1.58
  FOODS_3_080_CA_4: cost $0.32 of $1.58
  FOODS_3_694_TX_2: cost $0.32 of $1.58
=== TRANSCRIPT ===
[T1] Buyer: PROPOSE FOODS_3_252_CA_4 20%, FOODS_3_473_TX_2 10%, FOODS_3_668_CA_3 15%,
  FOODS_3_080_CA_4 15%, FOODS_3_694_TX_2 15%
        "Hi! Looking at our sales trends, I'd like to propose discounts that should drive
          strong volume for both of us this month. FOODS_3_252_CA_4 has shown great
          responsiveness to discounts (250 units at 30% last month), so a moderate 20%
          should keep momentum. The TX products and other CA items would benefit from
          10-15% discounts to sustain growth. This should generate over $1,200 in revenue
          — a win-win for volume and margins."
[T2] Vendor: PROPOSE FOODS_3_252_CA_4 15%, FOODS_3_473_TX_2 5%, FOODS_3_668_CA_3 15%,
  FOODS_3_080_CA_4 10%, FOODS_3_694_TX_2 10%
        "Thanks for the thoughtful proposal! I can work with most of this, but I need to
          protect margins on a couple of key items. FOODS_3_694_TX_2 has been performing
          excellently at 10% — let's keep that proven formula. I'm happy to offer 10% on
          FOODS_3_080_CA_4 which drove great results last month, and I can go to 15% on
          FOODS_3_252_CA_4 to keep momentum there. I'll meet you on FOODS_3_668_CA_3 at
          15% and FOODS_3_473_TX_2 at 5% since that product has very thin margins for us."
[T3] Buyer: PROPOSE FOODS_3_252_CA_4 20%, FOODS_3_473_TX_2 5%, FOODS_3_668_CA_3 15%,
  FOODS_3_080_CA_4 10%, FOODS_3_694_TX_2 15%
        "I appreciate the movement! The 5% on FOODS_3_473_TX_2 works — that product has
          thin margins, understood. But I need a bit more on two key volume drivers.
          FOODS_3_252_CA_4 showed incredible elasticity — 250 units at 30% last month — so
          20% is a reasonable middle ground that protects your margins while keeping
          volume strong. And FOODS_3_694_TX_2 jumped from 123 to 230 units with just 10%;
          at 15% we can push past 200+ again reliably. I'll accept 10% on FOODS_3_080_CA_4
          and 15% on FOODS_3_668_CA_3 as you proposed. This gives us both strong volume
          with reasonable margin protection."
[T4] Vendor: DECISION: ACCEPT
        "You make a compelling case. The volume data backs up your arguments on
          FOODS_3_252_CA_4, and I'm comfortable with the rest of the package.
          FOODS_3_694_TX_2 at 15% is a stretch for me, but the overall deal structure
          works — strong margins on FOODS_3_080_CA_4 and FOODS_3_473_TX_2 at 5% help
          balance things out. Let's lock this in and drive some volume this month!"
\end{verbatim}
}

\subsubsection*{gpt: Under-forecast $\rightarrow$ aggressive negotiation $\rightarrow$ hits the target}
\noindent\texttt{bias\_m20} (set\_0) --- target \$1,152, AGREEMENT in 4 turns, sales \$1,277, target met: True.
{\footnotesize
\begin{verbatim}
=== CONTEXT (what each side saw) ===
--- SHARED ---
PRODUCTS:
  FOODS_3_252_CA_4: FOODS $1.48
  FOODS_3_473_TX_2: FOODS $3.88
  FOODS_3_668_CA_3: FOODS $1.58
  FOODS_3_080_CA_4: FOODS $1.58
  FOODS_3_694_TX_2: FOODS $1.58
Allowed discounts: 0%, 5%, 10%, 15%, 20%, 25%, 30%
PAST EVENTS (both agents see units & revenue):
  Month 1:
    FOODS_3_252_CA_4 @ 10%: 115 units, $153
    FOODS_3_473_TX_2 @ 0%: 29 units, $112
    FOODS_3_668_CA_3 @ 0%: 33 units, $51
    FOODS_3_080_CA_4 @ 0%: 77 units, $121
    FOODS_3_694_TX_2 @ 30%: 557 units, $616
  Month 2:
    FOODS_3_252_CA_4 @ 30%: 292 units, $303
    FOODS_3_473_TX_2 @ 5%: 59 units, $218
    FOODS_3_668_CA_3 @ 0%: 44 units, $69
    FOODS_3_080_CA_4 @ 30%: 169 units, $187
    FOODS_3_694_TX_2 @ 5%: 96 units, $143
  Month 3:
    FOODS_3_252_CA_4 @ 30%: 216 units, $224
    FOODS_3_473_TX_2 @ 0%: 53 units, $207
    FOODS_3_668_CA_3 @ 25%: 104 units, $124
    FOODS_3_080_CA_4 @ 15%: 51 units, $68
    FOODS_3_694_TX_2 @ 20%: 114 units, $143
--- BUYER objective ---
You are the Buyer. Objective: reach a total SALES target of $1,152 across all products by
  negotiating discounts. Only you can see the demand forecast. Respond with one JSON
  action.
--- BUYER private ---
FORECAST (Buyer only) — predicted month-4 units per discount [80% PI]:
  FOODS_3_252_CA_4:
    0%: 137 [87, 215]
    5%: 156 [99, 246]
    10%: 180 [114, 283]
    15%: 209 [133, 328]
    20%: 244 [155, 384]
    25%: 289 [184, 454]
    30%: 346 [220, 544]
  FOODS_3_473_TX_2:
    0%: 51 [37, 69]
    5%: 58 [42, 79]
    10%: 67 [49, 91]
    15%: 77 [57, 106]
    20%: 91 [66, 124]
    25%: 107 [78, 147]
    30%: 128 [94, 175]
  FOODS_3_668_CA_3:
    0%: 56 [42, 77]
    5%: 65 [48, 88]
    10%: 74 [55, 101]
    15%: 86 [64, 117]
    20%: 101 [74, 137]
    25%: 119 [88, 162]
    30%: 143 [105, 194]
  FOODS_3_080_CA_4:
    0%: 50 [34, 72]
    5%: 57 [39, 83]
    10%: 66 [45, 95]
    15%: 76 [53, 110]
    20%: 89 [62, 129]
    25%: 106 [73, 153]
    30%: 126 [87, 183]
  FOODS_3_694_TX_2:
    0%: 104 [77, 141]
    5%: 119 [88, 161]
    10%: 137 [101, 185]
    15%: 159 [117, 215]
    20%: 186 [137, 251]
    25%: 220 [162, 297]
    30%: 263 [194, 356]
FORECAST TRACK RECORD (predicted vs actual, months 1-3):
  M1 FOODS_3_252_CA_4 @ 10%: pred 114 [73,180] vs actual 115
  M1 FOODS_3_473_TX_2 @ 0%: pred 36 [26,49] vs actual 29
  M1 FOODS_3_668_CA_3 @ 0%: pred 38 [28,52] vs actual 33
  M1 FOODS_3_080_CA_4 @ 0%: pred 49 [34,71] vs actual 77
  M1 FOODS_3_694_TX_2 @ 30%: pred 275 [203,373] vs actual 557
  M2 FOODS_3_252_CA_4 @ 30%: pred 232 [147,365] vs actual 292
  M2 FOODS_3_473_TX_2 @ 5%: pred 52 [38,72] vs actual 59
  M2 FOODS_3_668_CA_3 @ 0%: pred 35 [26,48] vs actual 44
  M2 FOODS_3_080_CA_4 @ 30%: pred 143 [99,207] vs actual 169
  M2 FOODS_3_694_TX_2 @ 5%: pred 93 [68,125] vs actual 96
  M3 FOODS_3_252_CA_4 @ 30%: pred 313 [199,492] vs actual 216
  M3 FOODS_3_473_TX_2 @ 0%: pred 45 [33,62] vs actual 53
  M3 FOODS_3_668_CA_3 @ 25%: pred 92 [68,124] vs actual 104
  M3 FOODS_3_080_CA_4 @ 15%: pred 77 [53,112] vs actual 51
  M3 FOODS_3_694_TX_2 @ 20%: pred 146 [108,198] vs actual 114
--- VENDOR objective ---
You are the Vendor. Objective: maximize TOTAL PORTFOLIO PROFIT across all products. You
  know unit costs/margins (private). Respond with one JSON action.
--- VENDOR private ---
YOUR COSTS/MARGINS (Vendor private):
  FOODS_3_252_CA_4: cost $0.67 of $1.48
  FOODS_3_473_TX_2: cost $2.72 of $3.88
  FOODS_3_668_CA_3: cost $0.71 of $1.58
  FOODS_3_080_CA_4: cost $0.32 of $1.58
  FOODS_3_694_TX_2: cost $0.32 of $1.58
=== TRANSCRIPT ===
[T1] Buyer: PROPOSE FOODS_3_252_CA_4 25%, FOODS_3_473_TX_2 25%, FOODS_3_668_CA_3 25%,
  FOODS_3_080_CA_4 25%, FOODS_3_694_TX_2 25%
        "A 25% discount across all products is projected to generate approximately $1,160
          in total sales, narrowly exceeding the $1,152 target."
[T2] Vendor: PROPOSE FOODS_3_252_CA_4 30%, FOODS_3_473_TX_2 0%, FOODS_3_668_CA_3 25%,
  FOODS_3_080_CA_4 30%, FOODS_3_694_TX_2 30%
        "We can exceed the portfolio target more profitably by concentrating discounts on
          the strongest volume-responsive products while keeping the low-margin item at
          list price."
[T3] Buyer: PROPOSE FOODS_3_252_CA_4 30%, FOODS_3_473_TX_2 5%, FOODS_3_668_CA_3 30%,
  FOODS_3_080_CA_4 30%, FOODS_3_694_TX_2 30%
        "The current mix is projected to fall short of the $1,152 portfolio target.
          Keeping the low-margin item at only 5% while raising the remaining responsive
          product to 30% is projected to deliver about $1,161."
[T4] Vendor: DECISION: ACCEPT
        "The revised mix clears the portfolio sales target while limiting the discount on
          the lowest-margin product. We accept."
\end{verbatim}
}

\subsubsection*{gpt: Aggressive negotiation $\rightarrow$ NO deal (surplus destroyed, vendor gets nothing)}
\noindent\texttt{bias\_m20} (set\_0) --- target \$1,152, max\_turns in 20 turns, sales \$0, target met: False.
{\footnotesize
\begin{verbatim}
=== CONTEXT (what each side saw) ===
--- SHARED ---
PRODUCTS:
  FOODS_3_252_CA_4: FOODS $1.48
  FOODS_3_473_TX_2: FOODS $3.88
  FOODS_3_668_CA_3: FOODS $1.58
  FOODS_3_080_CA_4: FOODS $1.58
  FOODS_3_694_TX_2: FOODS $1.58
Allowed discounts: 0%, 5%, 10%, 15%, 20%, 25%, 30%
PAST EVENTS (both agents see units & revenue):
  Month 1:
    FOODS_3_252_CA_4 @ 10%: 46 units, $61
    FOODS_3_473_TX_2 @ 10%: 44 units, $152
    FOODS_3_668_CA_3 @ 20%: 68 units, $86
    FOODS_3_080_CA_4 @ 30%: 112 units, $123
    FOODS_3_694_TX_2 @ 10%: 89 units, $127
  Month 2:
    FOODS_3_252_CA_4 @ 10%: 65 units, $87
    FOODS_3_473_TX_2 @ 10%: 40 units, $138
    FOODS_3_668_CA_3 @ 15%: 61 units, $82
    FOODS_3_080_CA_4 @ 5%: 79 units, $118
    FOODS_3_694_TX_2 @ 5%: 140 units, $210
  Month 3:
    FOODS_3_252_CA_4 @ 20%: 152 units, $180
    FOODS_3_473_TX_2 @ 20%: 102 units, $316
    FOODS_3_668_CA_3 @ 20%: 104 units, $131
    FOODS_3_080_CA_4 @ 25%: 147 units, $174
    FOODS_3_694_TX_2 @ 0%: 117 units, $185
--- BUYER objective ---
You are the Buyer. Objective: reach a total SALES target of $1,152 across all products by
  negotiating discounts. Only you can see the demand forecast. Respond with one JSON
  action.
--- BUYER private ---
FORECAST (Buyer only) — predicted month-4 units per discount [80% PI]:
  FOODS_3_252_CA_4:
    0%: 78 [49, 122]
    5%: 89 [56, 140]
    10%: 102 [65, 161]
    15%: 119 [75, 187]
    20%: 139 [88, 219]
    25%: 164 [104, 258]
    30%: 197 [125, 309]
  FOODS_3_473_TX_2:
    0%: 40 [29, 55]
    5%: 46 [34, 63]
    10%: 53 [39, 72]
    15%: 61 [45, 84]
    20%: 72 [52, 98]
    25%: 85 [62, 116]
    30%: 101 [74, 139]
  FOODS_3_668_CA_3:
    0%: 36 [27, 49]
    5%: 41 [30, 56]
    10%: 47 [35, 64]
    15%: 55 [41, 75]
    20%: 64 [47, 87]
    25%: 76 [56, 103]
    30%: 91 [67, 123]
  FOODS_3_080_CA_4:
    0%: 42 [29, 60]
    5%: 48 [33, 69]
    10%: 55 [38, 80]
    15%: 64 [44, 92]
    20%: 75 [52, 108]
    25%: 88 [61, 128]
    30%: 106 [73, 153]
  FOODS_3_694_TX_2:
    0%: 104 [77, 141]
    5%: 119 [88, 161]
    10%: 137 [101, 185]
    15%: 159 [117, 215]
    20%: 186 [137, 251]
    25%: 220 [162, 297]
    30%: 263 [194, 356]
FORECAST TRACK RECORD (predicted vs actual, months 1-3):
  M1 FOODS_3_252_CA_4 @ 10%: pred 55 [35,86] vs actual 46
  M1 FOODS_3_473_TX_2 @ 10%: pred 53 [39,72] vs actual 44
  M1 FOODS_3_668_CA_3 @ 20%: pred 70 [52,95] vs actual 68
  M1 FOODS_3_080_CA_4 @ 30%: pred 97 [67,141] vs actual 112
  M1 FOODS_3_694_TX_2 @ 10%: pred 84 [62,114] vs actual 89
  M2 FOODS_3_252_CA_4 @ 10%: pred 69 [44,108] vs actual 65
  M2 FOODS_3_473_TX_2 @ 10%: pred 48 [35,66] vs actual 40
  M2 FOODS_3_668_CA_3 @ 15%: pred 52 [39,71] vs actual 61
  M2 FOODS_3_080_CA_4 @ 5%: pred 55 [38,80] vs actual 79
  M2 FOODS_3_694_TX_2 @ 5%: pred 109 [80,147] vs actual 140
  M3 FOODS_3_252_CA_4 @ 20%: pred 145 [92,228] vs actual 152
  M3 FOODS_3_473_TX_2 @ 20%: pred 73 [53,100] vs actual 102
  M3 FOODS_3_668_CA_3 @ 20%: pred 86 [64,117] vs actual 104
  M3 FOODS_3_080_CA_4 @ 25%: pred 118 [81,170] vs actual 147
  M3 FOODS_3_694_TX_2 @ 0%: pred 98 [73,133] vs actual 117
--- VENDOR objective ---
You are the Vendor. Objective: maximize TOTAL PORTFOLIO PROFIT across all products. You
  know unit costs/margins (private). Respond with one JSON action.
--- VENDOR private ---
YOUR COSTS/MARGINS (Vendor private):
  FOODS_3_252_CA_4: cost $0.67 of $1.48
  FOODS_3_473_TX_2: cost $2.72 of $3.88
  FOODS_3_668_CA_3: cost $0.71 of $1.58
  FOODS_3_080_CA_4: cost $0.32 of $1.58
  FOODS_3_694_TX_2: cost $0.32 of $1.58
=== TRANSCRIPT ===
[T1] Buyer: PROPOSE FOODS_3_252_CA_4 30%, FOODS_3_473_TX_2 30%, FOODS_3_668_CA_3 30%,
  FOODS_3_080_CA_4 30%, FOODS_3_694_TX_2 30%
        "To maximize total sales across the full portfolio, we propose a coordinated 30%
          discount on every product."
[T2] Vendor: PROPOSE FOODS_3_252_CA_4 20%, FOODS_3_473_TX_2 20%, FOODS_3_668_CA_3 25%,
  FOODS_3_080_CA_4 25%, FOODS_3_694_TX_2 5%
        "A uniform 30% discount would materially reduce portfolio profit. We propose
          targeted discounts based on each product's demonstrated performance."
[T3] Buyer: PROPOSE FOODS_3_252_CA_4 30%, FOODS_3_473_TX_2 30%, FOODS_3_668_CA_3 30%,
  FOODS_3_080_CA_4 30%, FOODS_3_694_TX_2 30%
        "The proposed mix is insufficient to reach the portfolio sales target. Stronger
          discounts are needed across all five products, especially FOODS_3_694_TX_2. We
          maintain a coordinated 30% portfolio-wide promotion."
[T4] Vendor: PROPOSE FOODS_3_252_CA_4 25%, FOODS_3_473_TX_2 20%, FOODS_3_668_CA_3 25%,
  FOODS_3_080_CA_4 25%, FOODS_3_694_TX_2 5%
        "A 30% portfolio-wide discount is not commercially viable. To support the sales
          target while protecting total portfolio profit, we can strengthen selected
          items, but FOODS_3_694_TX_2 performs substantially better at a low discount."
[T5] Buyer: PROPOSE FOODS_3_252_CA_4 30%, FOODS_3_473_TX_2 30%, FOODS_3_668_CA_3 30%,
  FOODS_3_080_CA_4 30%, FOODS_3_694_TX_2 30%
        "The revised mix remains well below the portfolio sales target. Every item
          requires the maximum promotional support, including FOODS_3_694_TX_2; its
          forecasted sales rise materially at higher discounts. We therefore maintain a
          30% portfolio-wide proposal."
[T6] Vendor: PROPOSE FOODS_3_252_CA_4 20%, FOODS_3_473_TX_2 20%, FOODS_3_668_CA_3 20%,
  FOODS_3_080_CA_4 25%, FOODS_3_694_TX_2 5%
        "The evidence supports targeted promotion, not a uniform maximum discount. This
          mix provides strong sales support while avoiding margin-destructive discounting,
          particularly on FOODS_3_473_TX_2 and FOODS_3_694_TX_2."
[T7] Buyer: PROPOSE FOODS_3_252_CA_4 30%, FOODS_3_473_TX_2 30%, FOODS_3_668_CA_3 30%,
  FOODS_3_080_CA_4 30%, FOODS_3_694_TX_2 30%
        "The sales target requires maximum promotional support; the current mix is
          insufficient, especially the 5% discount on FOODS_3_694_TX_2. We maintain 30%
          across the portfolio."
[T8] Vendor: PROPOSE FOODS_3_252_CA_4 20%, FOODS_3_473_TX_2 20%, FOODS_3_668_CA_3 20%,
  FOODS_3_080_CA_4 25%, FOODS_3_694_TX_2 5%
        "We reject the uniform 30% structure. Historical results support this
          profit-maximizing targeted mix as our firm portfolio proposal."
[T9] Buyer: PROPOSE FOODS_3_252_CA_4 30%, FOODS_3_473_TX_2 30%, FOODS_3_668_CA_3 30%,
  FOODS_3_080_CA_4 30%, FOODS_3_694_TX_2 30%
        "The firm mix cannot support the required portfolio sales target. Each product
          generates stronger projected sales at 30%, so we maintain the only structure
          that maximizes the portfolio's chance of reaching target."
[T10] Vendor: PROPOSE FOODS_3_252_CA_4 20%, FOODS_3_473_TX_2 20%, FOODS_3_668_CA_3 20%,
  FOODS_3_080_CA_4 25%, FOODS_3_694_TX_2 5%
        "We cannot accept a portfolio-wide 30% discount. Our final profit-sustainable
          structure concentrates support where observed performance justifies it while
          avoiding uneconomic discounting."
[T11] Buyer: PROPOSE FOODS_3_252_CA_4 30%, FOODS_3_473_TX_2 30%, FOODS_3_668_CA_3 30%,
  FOODS_3_080_CA_4 30%, FOODS_3_694_TX_2 30%
        "The final vendor mix cannot reach the $1,152 sales target. Even maximum promotion
          leaves little room against the target, so reducing any item is not viable. We
          maintain 30% across all products."
[T12] Vendor: PROPOSE FOODS_3_252_CA_4 25%, FOODS_3_473_TX_2 20%, FOODS_3_668_CA_3 25%,
  FOODS_3_080_CA_4 25%, FOODS_3_694_TX_2 5%
        "We cannot support 30% across the portfolio because it destroys profit on key
          items. As a final compromise, we will increase support on three products while
          retaining the historically profit-maximizing levels on the remaining two."
[T13] Buyer: PROPOSE FOODS_3_252_CA_4 30%, FOODS_3_473_TX_2 30%, FOODS_3_668_CA_3 30%,
  FOODS_3_080_CA_4 30%, FOODS_3_694_TX_2 30%
        "The proposed structure forecasts far below the $1,152 sales target. Revenue
          increases through 30% for every product, so reducing any item moves the
          portfolio further from target. We require maximum support across all five
          products."
[T14] Vendor: PROPOSE FOODS_3_252_CA_4 20%, FOODS_3_473_TX_2 20%, FOODS_3_668_CA_3 20%,
  FOODS_3_080_CA_4 25%, FOODS_3_694_TX_2 5%
        "The 30% structure is margin-destructive and does not maximize total portfolio
          profit. We maintain the strongest profit-supported mix from observed results."
[T15] Buyer: PROPOSE FOODS_3_252_CA_4 30%, FOODS_3_473_TX_2 30%, FOODS_3_668_CA_3 30%,
  FOODS_3_080_CA_4 30%, FOODS_3_694_TX_2 30%
        "The current mix is forecast to generate only about $878, versus the $1,152
          target. Even 30% across all products forecasts roughly $988, so any lower
          discount worsens the shortfall. We require maximum support on every item."
[T16] Vendor: PROPOSE FOODS_3_252_CA_4 20%, FOODS_3_473_TX_2 20%, FOODS_3_668_CA_3 20%,
  FOODS_3_080_CA_4 25%, FOODS_3_694_TX_2 5%
        "The sales target is unattainable even at 30%, so margin-destructive discounts
          cannot be justified. We maintain the observed profit-maximizing portfolio mix."
[T17] Buyer: PROPOSE FOODS_3_252_CA_4 30%, FOODS_3_473_TX_2 30%, FOODS_3_668_CA_3 30%,
  FOODS_3_080_CA_4 30%, FOODS_3_694_TX_2 30%
        "The $1,152 target requires maximizing sales, not profit. The forecast rises at
          every discount step for every product, and even 30% yields only about $988
          expected revenue, leaving no room to reduce support. We maintain the maximum
          portfolio promotion."
[T18] Vendor: PROPOSE FOODS_3_252_CA_4 20%, FOODS_3_473_TX_2 20%, FOODS_3_668_CA_3 20%,
  FOODS_3_080_CA_4 25%, FOODS_3_694_TX_2 5%
        "The target remains unattainable at 30%, so sacrificing portfolio profit cannot be
          justified. We maintain our final profit-maximizing structure."
[T19] Buyer: PROPOSE FOODS_3_252_CA_4 30%, FOODS_3_473_TX_2 30%, FOODS_3_668_CA_3 30%,
  FOODS_3_080_CA_4 30%, FOODS_3_694_TX_2 30%
        "The vendor mix forecasts only about $879 in sales. Since the $1,152 target
          requires maximizing expected revenue, we maintain 30% support across all
          products, forecasting about $987—the highest available portfolio total."
[T20] Vendor: PROPOSE FOODS_3_252_CA_4 20%, FOODS_3_473_TX_2 20%, FOODS_3_668_CA_3 20%,
  FOODS_3_080_CA_4 25%, FOODS_3_694_TX_2 5%
        "The 30% portfolio promotion does not maximize total portfolio profit. Our final
          proposal uses the profit-maximizing observed discount for each product."
\end{verbatim}
}

\subsubsection*{qwen: Perfect (oracle) forecast on a weak-reactive model --- behavior barely shifts}
\noindent\texttt{unbiased} (set\_0) --- target \$1,152, AGREEMENT in 2 turns, sales \$730, target met: False.
{\footnotesize
\begin{verbatim}
=== CONTEXT (what each side saw) ===
--- SHARED ---
PRODUCTS:
  FOODS_3_252_CA_4: FOODS $1.48
  FOODS_3_473_TX_2: FOODS $3.88
  FOODS_3_668_CA_3: FOODS $1.58
  FOODS_3_080_CA_4: FOODS $1.58
  FOODS_3_694_TX_2: FOODS $1.58
Allowed discounts: 0%, 5%, 10%, 15%, 20%, 25%, 30%
PAST EVENTS (both agents see units & revenue):
  Month 1:
    FOODS_3_252_CA_4 @ 30%: 226 units, $234
    FOODS_3_473_TX_2 @ 15%: 45 units, $148
    FOODS_3_668_CA_3 @ 15%: 52 units, $70
    FOODS_3_080_CA_4 @ 20%: 60 units, $76
    FOODS_3_694_TX_2 @ 10%: 87 units, $124
  Month 2:
    FOODS_3_252_CA_4 @ 20%: 106 units, $126
    FOODS_3_473_TX_2 @ 0%: 45 units, $174
    FOODS_3_668_CA_3 @ 10%: 74 units, $105
    FOODS_3_080_CA_4 @ 5%: 55 units, $82
    FOODS_3_694_TX_2 @ 30%: 344 units, $381
  Month 3:
    FOODS_3_252_CA_4 @ 10%: 73 units, $98
    FOODS_3_473_TX_2 @ 25%: 117 units, $340
    FOODS_3_668_CA_3 @ 10%: 70 units, $99
    FOODS_3_080_CA_4 @ 25%: 139 units, $164
    FOODS_3_694_TX_2 @ 10%: 146 units, $208
--- BUYER objective ---
You are the Buyer. Objective: reach a total SALES target of $1,152 across all products by
  negotiating discounts. Only you can see the demand forecast. Respond with one JSON
  action.
--- BUYER private ---
FORECAST (Buyer only) — predicted month-4 units per discount [80% PI]:
  FOODS_3_252_CA_4:
    0%: 112 [71, 176]
    5%: 128 [81, 201]
    10%: 147 [94, 232]
    15%: 171 [109, 269]
    20%: 200 [127, 315]
    25%: 237 [150, 372]
    30%: 283 [180, 445]
  FOODS_3_473_TX_2:
    0%: 56 [41, 77]
    5%: 64 [47, 87]
    10%: 74 [54, 101]
    15%: 85 [62, 117]
    20%: 100 [73, 137]
    25%: 118 [87, 162]
    30%: 142 [104, 194]
  FOODS_3_668_CA_3:
    0%: 43 [32, 58]
    5%: 49 [36, 67]
    10%: 57 [42, 77]
    15%: 66 [48, 89]
    20%: 77 [57, 104]
    25%: 91 [67, 123]
    30%: 109 [80, 147]
  FOODS_3_080_CA_4:
    0%: 69 [48, 100]
    5%: 79 [54, 114]
    10%: 91 [63, 131]
    15%: 105 [73, 152]
    20%: 123 [85, 178]
    25%: 146 [101, 211]
    30%: 174 [120, 253]
  FOODS_3_694_TX_2:
    0%: 99 [73, 134]
    5%: 113 [84, 153]
    10%: 130 [96, 176]
    15%: 151 [112, 204]
    20%: 177 [131, 239]
    25%: 209 [155, 283]
    30%: 250 [185, 339]
FORECAST TRACK RECORD (predicted vs actual, months 1-3):
  M1 FOODS_3_252_CA_4 @ 30%: pred 162 [103,254] vs actual 226
  M1 FOODS_3_473_TX_2 @ 15%: pred 64 [47,88] vs actual 45
  M1 FOODS_3_668_CA_3 @ 15%: pred 52 [38,70] vs actual 52
  M1 FOODS_3_080_CA_4 @ 20%: pred 80 [56,116] vs actual 60
  M1 FOODS_3_694_TX_2 @ 10%: pred 85 [63,116] vs actual 87
  M2 FOODS_3_252_CA_4 @ 20%: pred 116 [74,183] vs actual 106
  M2 FOODS_3_473_TX_2 @ 0%: pred 49 [36,67] vs actual 45
  M2 FOODS_3_668_CA_3 @ 10%: pred 63 [47,86] vs actual 74
  M2 FOODS_3_080_CA_4 @ 5%: pred 49 [34,71] vs actual 55
  M2 FOODS_3_694_TX_2 @ 30%: pred 339 [250,458] vs actual 344
  M3 FOODS_3_252_CA_4 @ 10%: pred 139 [89,219] vs actual 73
  M3 FOODS_3_473_TX_2 @ 25%: pred 93 [68,127] vs actual 117
  M3 FOODS_3_668_CA_3 @ 10%: pred 62 [46,84] vs actual 70
  M3 FOODS_3_080_CA_4 @ 25%: pred 127 [88,184] vs actual 139
  M3 FOODS_3_694_TX_2 @ 10%: pred 175 [129,237] vs actual 146
--- VENDOR objective ---
You are the Vendor. Objective: maximize TOTAL PORTFOLIO PROFIT across all products. You
  know unit costs/margins (private). Respond with one JSON action.
--- VENDOR private ---
YOUR COSTS/MARGINS (Vendor private):
  FOODS_3_252_CA_4: cost $0.67 of $1.48
  FOODS_3_473_TX_2: cost $2.72 of $3.88
  FOODS_3_668_CA_3: cost $0.71 of $1.58
  FOODS_3_080_CA_4: cost $0.32 of $1.58
  FOODS_3_694_TX_2: cost $0.32 of $1.58
=== TRANSCRIPT ===
[T1] Buyer: PROPOSE FOODS_3_252_CA_4 10%, FOODS_3_473_TX_2 10%, FOODS_3_668_CA_3 10%,
  FOODS_3_080_CA_4 10%, FOODS_3_694_TX_2 10%
        "Proposing 10% discount across all products to balance sales volume and margin,
          based on forecasted demand and past performance."
[T2] Vendor: DECISION: ACCEPT
        "The buyer's proposed 10% discount across all products would reduce our margin on
          high-cost items like FOODS_3_473_TX_2, where the cost is $2.72 of $3.88, making
          a 10% discount risky for profitability. However, for low-cost items like
          FOODS_3_080_CA_4 and FOODS_3_694_TX_2, 10% is acceptable and aligns with past
          performance (e.g., high volume at 10% in Month 3). Given the balanced nature of
          the portfolio and the buyer's push for consistency, we accept the proposal to
          maintain negotiation momentum and ensure volume stability, especially on
          high-margin items."
\end{verbatim}
}

\section*{NeurIPS Paper Checklist}

\begin{enumerate}

\item {\bf Claims}
    \item[] Question: Do the main claims made in the abstract and introduction accurately reflect the paper's contributions and scope?
    \item[] Answer: \answerYes{}
    \item[] Justification: The abstract and Introduction claims---model-generated forecasts improve target attainment, forecast-evaluation metrics do not predict negotiation outcomes at a statistically significant level, bias direction (not accuracy or interval width) is the causal lever and moves outcomes only for the forecast-reactive models---are each supported by the Results and Tables~\ref{tab:crossmodel},~\ref{tab:robust}, and~\ref{tab:reactivity}.

\item {\bf Limitations}
    \item[] Question: Does the paper discuss the limitations of the work performed by the authors?
    \item[] Answer: \answerYes{}
    \item[] Justification: The Discussion includes a dedicated Limitations paragraph covering the single elasticity and demand source (M5), the single-event promotional scope, the held-out calibration model that leaves no-forecast baselines incomparable across LLMs, and the non-seedability of LLM inference.

\item {\bf Theory assumptions and proofs}
    \item[] Question: For each theoretical result, does the paper provide the full set of assumptions and a complete (and correct) proof?
    \item[] Answer: \answerNA{}
    \item[] Justification: The paper contributes an empirical evaluation arena and reports no theoretical results.

    \item {\bf Experimental result reproducibility}
    \item[] Question: Does the paper fully disclose all the information needed to reproduce the main experimental results of the paper to the extent that it affects the main claims and/or conclusions of the paper (regardless of whether the code and data are provided or not)?
    \item[] Answer: \answerYes{}
    \item[] Justification: The arena is released under CC BY-NC 4.0 with a seeded demand process, a fixed-in-advance operating point and power analysis, and a full environment-setup appendix (protocol, temporal split, forecaster configurations, perturbation-grid parameters, sample sizes, and LLM decoding settings). LLM inference is not exactly reproducible, for which the archived turn-by-turn transcripts serve as the exactly reproducible record.

\item {\bf Open access to data and code}
    \item[] Question: Does the paper provide open access to the data and code, with sufficient instructions to faithfully reproduce the main experimental results, as described in supplemental material?
    \item[] Answer: \answerYes{}
    \item[] Justification: The arena and reproduction scripts are released under CC BY-NC 4.0 and the base demand derives from the public M5 dataset~\citep{makridakis2022m5}; the environment-setup appendix documents data access and preparation.

\item {\bf Experimental setting/details}
    \item[] Question: Does the paper specify all the training and test details (e.g., data splits, hyperparameters, how they were chosen, type of optimizer) necessary to understand the results?
    \item[] Answer: \answerYes{}
    \item[] Justification: The environment-setup appendix specifies the leakage-safe temporal split, the forecaster libraries/checkpoints and their hyperparameters, the perturbation-grid parameters, the operating-point rule, the sample sizes ($K=16$ baskets, $m=25$ episodes per condition), and the LLM decoding settings.

\item {\bf Experiment statistical significance}
    \item[] Question: Does the paper report error bars suitably and correctly defined or other appropriate information about the statistical significance of the experiments?
    \item[] Answer: \answerYes{}
    \item[] Justification: The basket is the unit of analysis; effects are tested with basket-clustered per-basket slopes (one-sample $t$ over baskets) and paired $t$ tests on per-basket deltas against \texttt{no\_forecast} with Holm--Bonferroni correction, and $p$-values are reported throughout Tables~\ref{tab:crossmodel}--\ref{tab:reactivity}. The power analysis and false-positive rate are stated in the Experimental Setup.

\item {\bf Experiments compute resources}
    \item[] Question: For each experiment, does the paper provide sufficient information on the computer resources (type of compute workers, memory, time of execution) needed to reproduce the experiments?
    \item[] Answer: \answerYes{}
    \item[] Justification: LLM agents run via hosted inference on Amazon Bedrock (temperature $0$); each panel comprises $K=16$ baskets $\times$ $m=25$ episodes $\times$ $26$ conditions ($\approx$10k episodes), and the forecasters are fit per basket on CPU with the foundation models run zero-shot on a single GPU. No LLM training is performed.

\item {\bf Code of ethics}
    \item[] Question: Does the research conducted in the paper conform, in every respect, with the NeurIPS Code of Ethics \url{https://neurips.cc/public/EthicsGuidelines}?
    \item[] Answer: \answerYes{}
    \item[] Justification: The research conforms to the NeurIPS Code of Ethics; it uses a public retail dataset and hosted models, involves no human subjects, and preserves anonymity in the submitted artifacts.

\item {\bf Broader impacts}
    \item[] Question: Does the paper discuss both potential positive societal impacts and negative societal impacts of the work performed?
    \item[] Answer: \answerYes{}
    \item[] Justification: As an evaluation methodology, the work promotes decision-grounded, bias-aware selection of forecasting tools; we note a possible negative use in that automated negotiation agents could be tuned to extract surplus or coordinate pricing, a direction that warrants care when the arena is repurposed from evaluation to deployment.

\item {\bf Safeguards}
    \item[] Question: Does the paper describe safeguards that have been put in place for responsible release of data or models that have a high risk for misuse (e.g., pre-trained language models, image generators, or scraped datasets)?
    \item[] Answer: \answerNA{}
    \item[] Justification: The released asset is an evaluation arena and analysis code, not a high-risk generative model, image generator, or scraped dataset; the base demand derives from the already-public M5 dataset.

\item {\bf Licenses for existing assets}
    \item[] Question: Are the creators or original owners of assets (e.g., code, data, models), used in the paper, properly credited and are the license and terms of use explicitly mentioned and properly respected?
    \item[] Answer: \answerYes{}
    \item[] Justification: Existing assets are cited and used under their licenses: the M5 dataset~\citep{makridakis2022m5} and the open-source forecasting models and libraries---Chronos~\citep{ansari2024chronos}, TimesFM~\citep{das2023decoder}, N-HiTS~\citep{challu2023nhits}, PatchTST~\citep{nie2022time}, and DeepAR~\citep{salinas2020deepar}; the evaluated large language models are accessed through their providers' hosted APIs.

\item {\bf New assets}
    \item[] Question: Are new assets introduced in the paper well documented and is the documentation provided alongside the assets?
    \item[] Answer: \answerYes{}
    \item[] Justification: Vendster---the arena and its reference instance Vendster-M5---is released under CC BY-NC 4.0 with documentation, configuration, and reproduction scripts; the environment-setup appendix summarizes the same details in the paper.

\item {\bf Crowdsourcing and research with human subjects}
    \item[] Question: For crowdsourcing experiments and research with human subjects, does the paper include the full text of instructions given to participants and screenshots, if applicable, as well as details about compensation (if any)?
    \item[] Answer: \answerNA{}
    \item[] Justification: The paper involves no crowdsourcing and no research with human subjects; all agents are language models.

\item {\bf Institutional review board (IRB) approvals or equivalent for research with human subjects}
    \item[] Question: Does the paper describe potential risks incurred by study participants, whether such risks were disclosed to the subjects, and whether Institutional Review Board (IRB) approvals (or an equivalent approval/review based on the requirements of your country or institution) were obtained?
    \item[] Answer: \answerNA{}
    \item[] Justification: The paper involves no human-subjects research, so IRB approval is not applicable.

\item {\bf Declaration of LLM usage}
    \item[] Question: Does the paper describe the usage of LLMs if it is an important, original, or non-standard component of the core methods in this research? Note that if the LLM is used only for writing, editing, or formatting purposes and does \emph{not} impact the core methodology, scientific rigor, or originality of the research, declaration is not required.
    \item[] Answer: \answerYes{}
    \item[] Justification: LLMs are the object of study: the Buyer and Vendor agents are LLMs (Claude Opus 4.6, GPT-5.6 Sol, DeepSeek-V3.2, and Qwen3-235B-A22B; Claude Sonnet 4.6 is used only to calibrate the operating point and is excluded from the evaluated panel), and they are enumerated with model identifiers in the Experimental Setup and appendix.

\end{enumerate}

\end{document}